\documentclass[aps,10pt,final,titlepage,oneside,twocolumn,nobibnotes,nofootinbib,superscriptaddress,noshowpacs,centertags]{revtex4}
\usepackage{amssymb,amsmath}
\usepackage{eufrak}
\usepackage{mathtext}
\usepackage[utf8]{inputenc}
\usepackage[T2A]{fontenc}
\usepackage[russian,english]{babel}
\usepackage{bm}
\usepackage{graphicx}
\usepackage{calc}
\usepackage{ifthen}
\usepackage{caption2}
\usepackage{epstopdf}
\usepackage{multirow}
\newcommand{\chem}[1]{$\mathop{\mathrm{#1}}$}
\begin{document}
\selectlanguage{english}

\title{The influence of anisotropy on the manifestation of aging effects in the magnetoresistance of multilayer structures}

\author{Marina M. Boldyreva$^{1,3}$, Dmitry A. Druzev$^1$, Vadim O. Borzilov$^1$, Igor K. Saifutdinov$^1$, Natalia I. Piskunova$^1$, Marina V. Mamonova$^1$, Vladimir V. Prudnikov}%
\affiliation{
Department of Theoretical Physics, Dostoevsky Omsk State University, Mira Av. 55-A, Omsk 644077, Russia.}
\author{Pavel V. Prudnikov}%
\affiliation{Center of New Chemical Technologies BIC, Boreskov Institute of Catalysis, Neftezavodskay str. 54, Omsk 644040, Russia.\\
$^3$Laboratory of Theoretical and Computational Nanoscience, Kakumamachi 920-1192, Kanazawa, Japan.}

E-mail: ffi-95@mail.ru \\

\begin{abstract}

Currently, aging effects occurring in multilayer systems attract much attention from researchers. In our research, we report the investigation on the
non-equilibrium relaxations of the ferromagnetic-nonmagnetic multilayer \chem{Co/Cu/Co} and \chem{Pt/Co/Cu/Co/Pt} by means of the Monte
Carlo simulation using that depend on anisotropy, magnetoresistance and susceptibility. All these parameters are interrelated and have varying
degrees of influence on the effects of aging. We researched our Co-layers magnetic system from high-temperature and low-temperature initial
states, as well as not only in the critical temperature region, but in a wide low-temperature range. The study was carried out with different types of
magnetic anisotropy, which made it possible to identify differences in the behavior of the system depending on the type of anisotropy. We found
that in systems exhibiting at dimensional transition, aging effects manifest to gain and themselves in a wide low-temperature range, and not only 
in a narrow near-critical region, which ultimately leads us to a wide range of applications of these materials for practical applications. 
We have conducted research on the non-equilibrium relaxations of the magnetic multipliers using the Monte Carlo simulation. The present study is located
within the series of the research.

\end{abstract}

\maketitle

\section{INTRODUCTION}

\begin{figure*}[ht!]
\centering
{\includegraphics[width=0.36\textwidth]{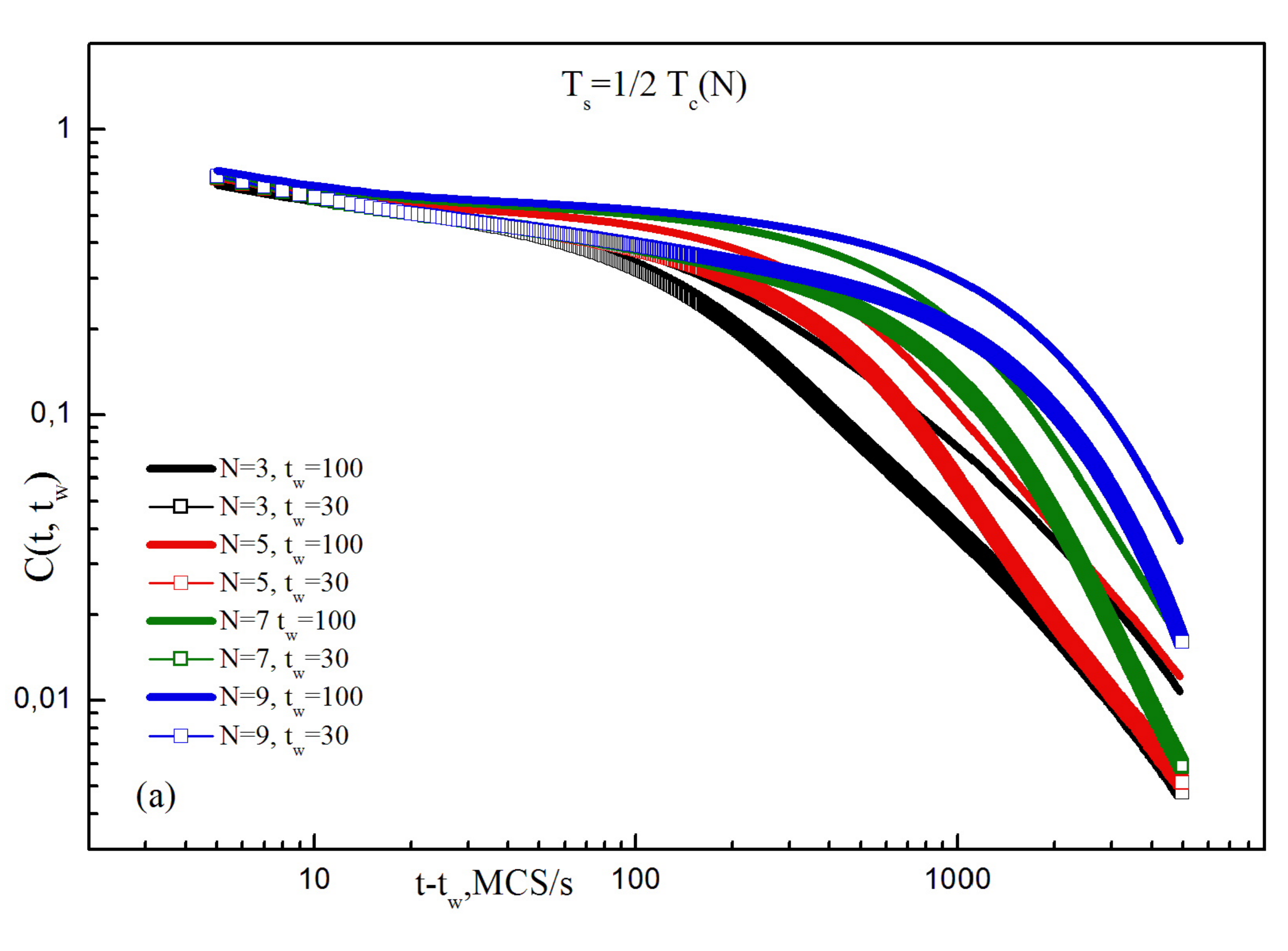}%
} \hfil
{\includegraphics[width=0.35\textwidth]{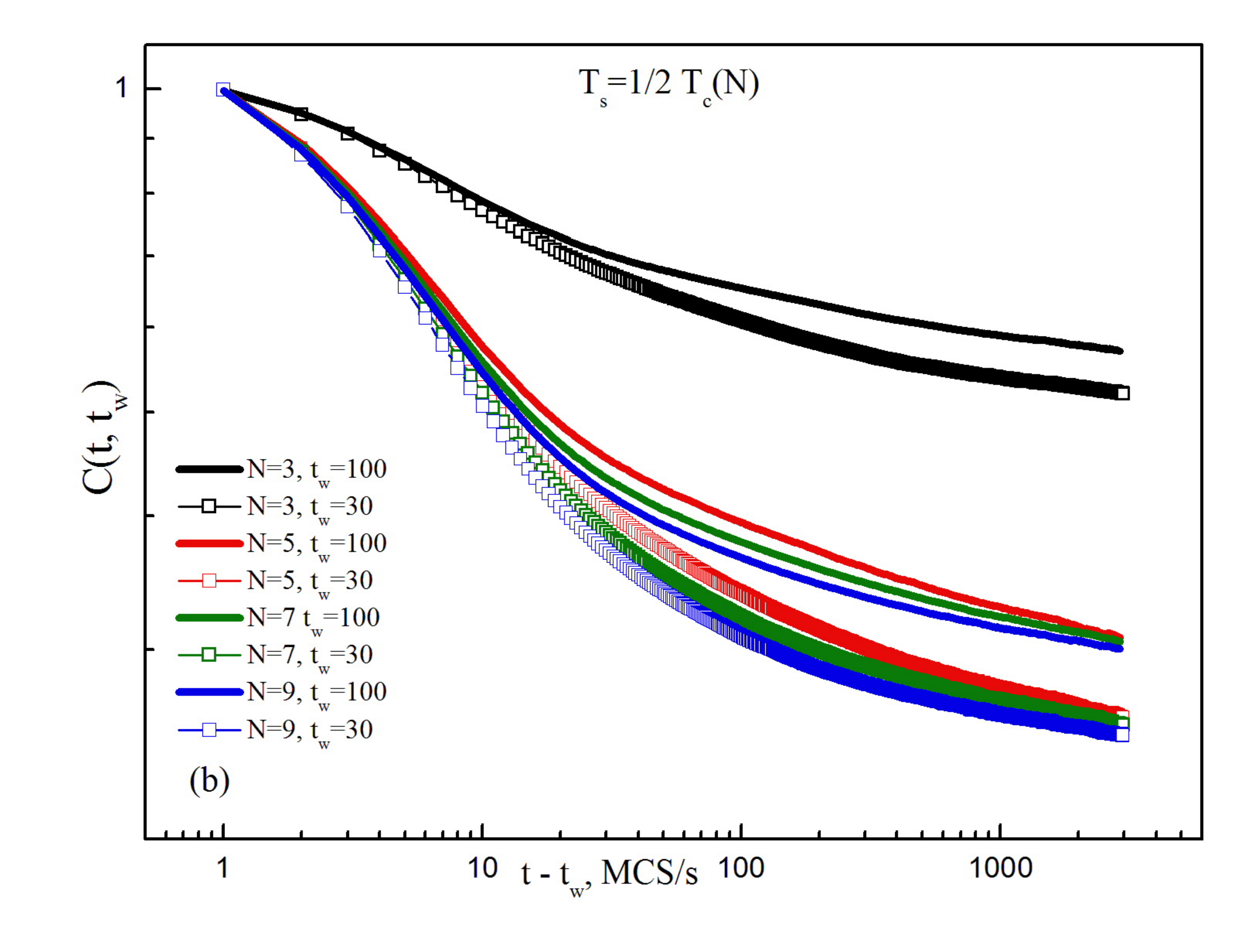}\\[-2mm]%
} \caption{\label{fig:image_1} The graphs show the dependency of $C(t,t_w)$ as a function $t-t_w$ where $t_w = 30$, $100$ MCS/s for \chem{Co/Cu(100)/Co}(a) and \chem{Pt/Co/Cu(100)/Co/Pt}(b) structures with \chem{Co} 
where film thicknesses took as $N = 3; 5; 7; 9$ ML at temperatures $T_c/2$. Shown here is evolution from high-temperature initial state.}
\end{figure*}

The physics of ultrathin films and multilayer structures has been an area of active exploration for many years \cite{Dieny2020, Vaz2008}. The
interest in multilayer structures is related to a number of uncommon properties of this type of films with a thickness of one to a few tens of
monolayers. Areas of research aimed at the occurrence of aging effects still have broad potential for research  \cite{Bohmer, arxiv_Baity}. Alloys of
ultrathin magnetic films are used in various fields ranging from spintronic devices to biomedicine. These alloys are widely used in various fields,
one of which is the use as parts of multilayer structures in devices based on a tunnel and giant magnetoresistance \cite{CHEN, DUINE}.

In multilayer structures, nanoscale periodicity affects the specific relationship between the spin dependence with the slow dynamics of
magnetization relaxation, which also leads to quenching of the system in the case of a nonequilibrium state of the system. Magnetic multilayer
systems with nanoscale periodicity show to increase provide relaxation time, which due to the effects associated with a larger characteristic length
of the spin-spin correlation. This is one of the differences multilayer systems compare from a bulk. In the case of bulk systems, the effects of slow
dynamics and aging observe near critical point. Magnetic aging systems based on \chem{Co/Cr} was discovered in experimental studies, the
relaxation other magnetization of the system \cite{Mukherjee2010}.

Firstly, I would like to note that the presented study is a continuation and development of a number of studies indicated in the article by the authors. 
In our opinion, this study is of great importance for a number of reasons, one of which is the possibility of practical application of the materials under 
study in various fields of spintronics. In the course of our research presented in the articles \cite{JETPL2016, MAIN}, we discovered aging effects in the behavior of multilayer systems
\chem{Co/Cr/Co} and \chem{Pt/Co/Cu/Co/Pt}. 

In this work, the main attention is paid to an in-depth study of the properties of multilayer structures based on \chem{Co} for systems with different 
types of magnetic anisotropy. When modeling a multilayer structure, we expect to detect aging effects on the behavior of a quantity such as magnetoresistance 
and autocorrelation function. It is planned to determine the influence of the initial states of the system on the behavior of the magnetoresistance. 
The present study is located within the series of the research. In the current work, we obtain new data and results on the behavior of the autocorrelation 
function and susceptibility and other physical quantities, about which more detailed information can be found in the following sections.

\section{DESCRIPTION OF THE MODEL AND METHODS}

To carry out the research, we used a model of a multilayer structure, in which the system consists of ferromagnetic films with a separating layer of
non-magnetic metal film. In the article \cite{MAIN} we presented the results of an analysis of the dependence of the autocorrelation function $C (t,
t_ {w}) $ on the observation time $t-t_ {w} $ and for various times of expectation, so we will not well on already obtained results, and in this article
we will consider new results.

\begin{figure*}[!t]
\centering
{\includegraphics[width=0.25\textwidth]{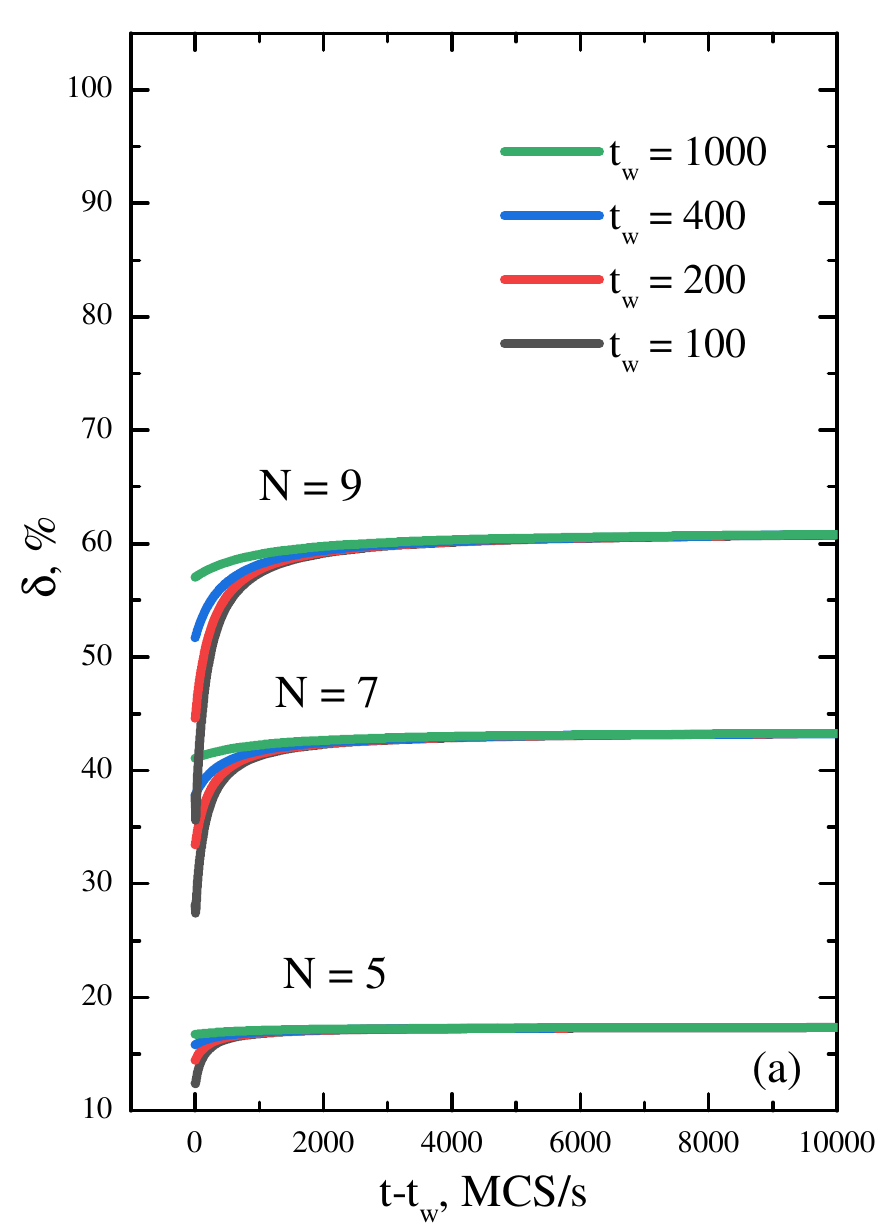}%
} \hfil
{\includegraphics[width=0.25\textwidth]{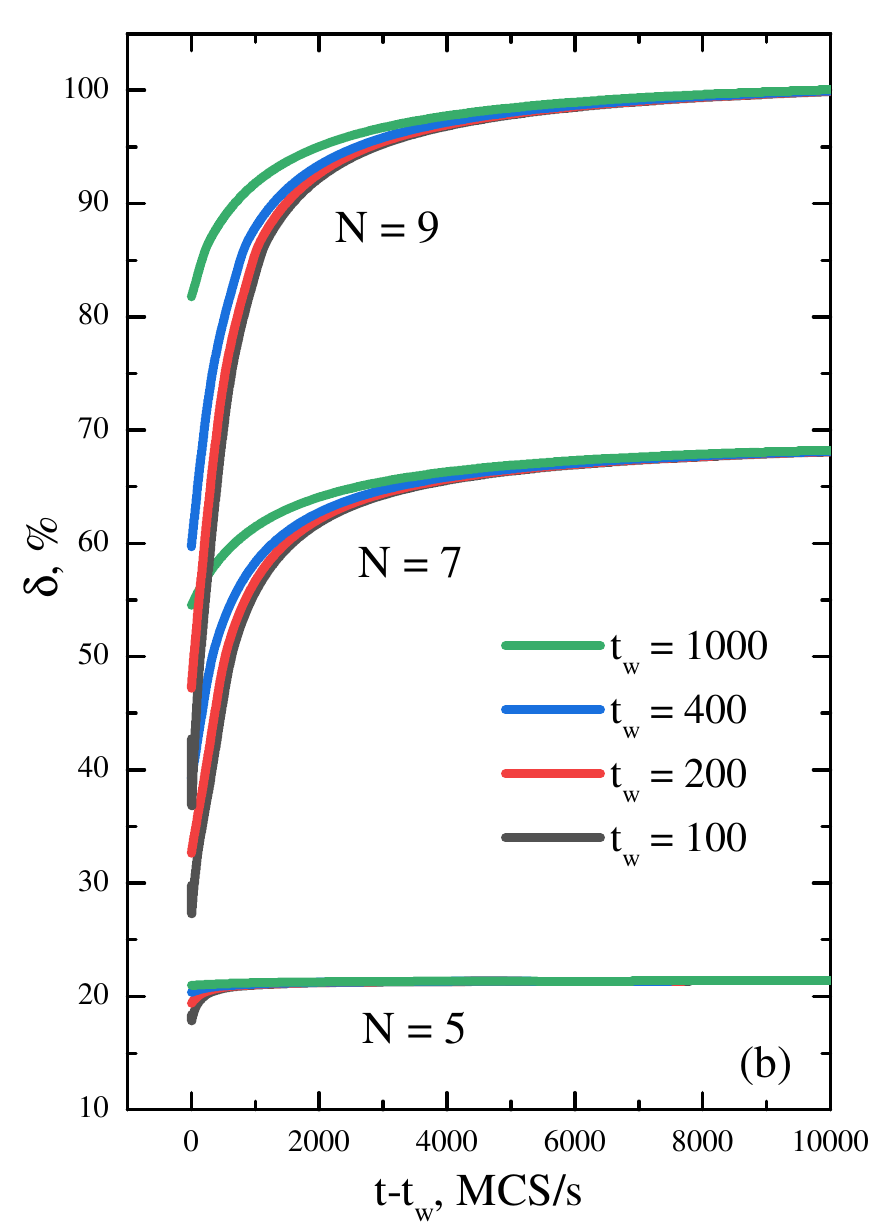}%
} \caption{\label{fig:image_5} The graphs show the time dependence of the CPP-magnetoresistance in \chem{Pt/Co/Cu(100)/Co/Pt} with the thicknesses $N = 5; 7; 9$ ML 
of the cobalt films at temperatures $T=T_c/4$ for different waiting times $t_w$ with evolution from the high-temperature (a) and from the low-temperature initial states (b).}
\end{figure*}

\begin{figure*}[!t]
\centering
{\includegraphics[width=0.25\textwidth]{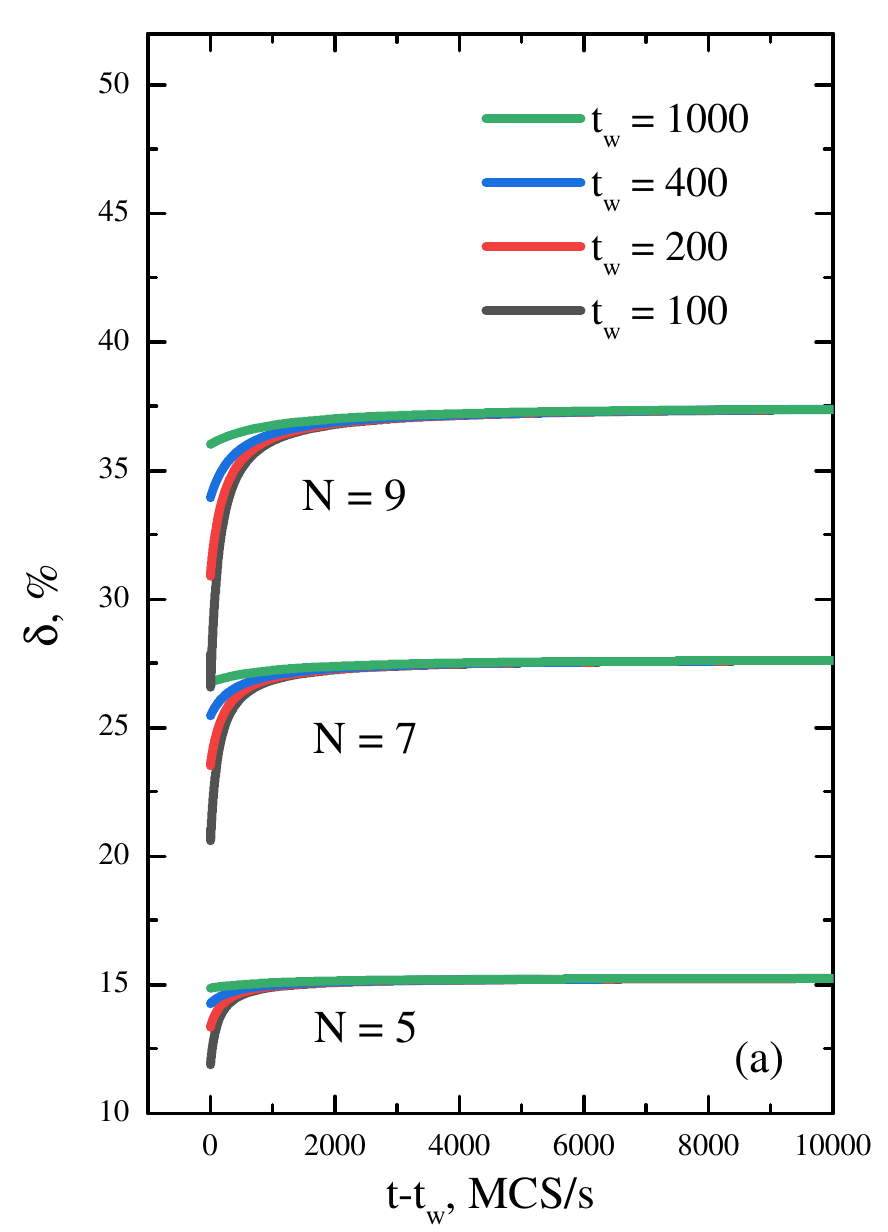}%
} \hfil
{\includegraphics[width=0.25\textwidth]{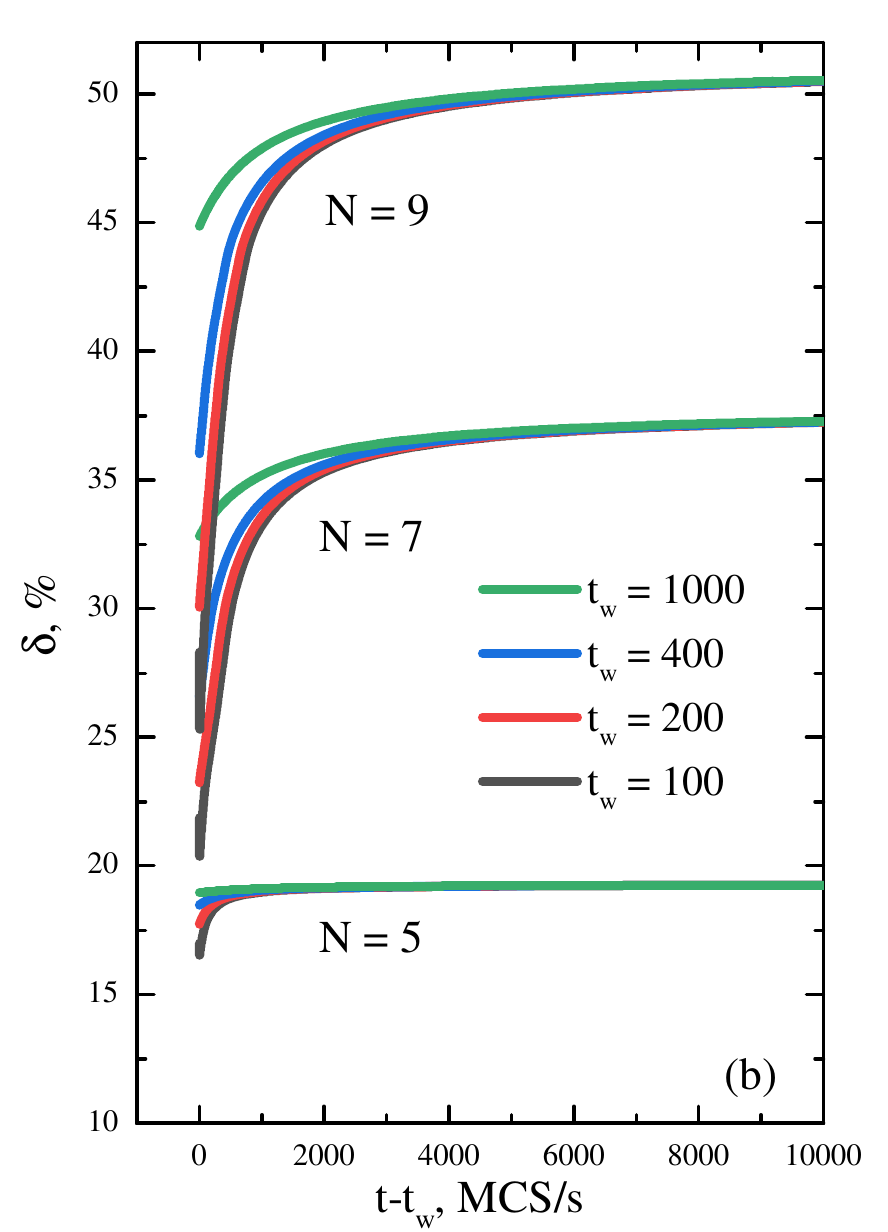}%
} \caption{\label{fig:image_7} The graphs show the time dependence of the CIP-magnetoresistance in \chem{Co/Cu(100)/Co}  with the thicknesses $N = 5; 7; 9$ ML 
of the cobalt films at temperatures $T=T_c/4$ for different waiting times $t_w$ with evolution from the high-temperature (a) and from the low-temperature initial states (b).}
\end{figure*}

Both in previous work and currently we are considering structures consisting of ferromagnetic films $N = 3; 5; 7; 9$ in units of monatomic layers 
(ML). The choice of thicknesses indicated in the study is due to the fact that with increasing film thickness at dimensional transition is observed. At
$N\leq 5$, the critical behavior corresponds to the 2d Ising model, and with increasing thickness $6 \leq N \leq 22$, a crossover transition from 2d
to 3d properties is observed \cite{JETPL2014}. This, the that attracts the most attention is $N=9$, that is, the case of system behavior during the
transition to three-dimensional properties.

The exchange integral, which determines the interaction of neighboring spins $J_1$, takes the value $J_1/k_BT=1$ and $J_2 = - 0.3 J_1$. Based
on some of the previously obtained result we came to the conclusion to continue the study of ultrathin magnetic films in a wider area, which
will be presented in our articles.

In the presented research, we simulated a multilayer magnetic system based on the Monte Carlo method to determine the nonequilibrium behavior
of the system. The Hamiltonian presented below was used for calculations
\begin{multline}
H=-\sum_{\langle i,j\rangle} J_{ij} \bigg\{ \left(S_{i}^{x} S_{j}^{x}+S_{i}^{y} S_{j}^{y}\right) \\ + \left( 1-\Delta_{1}(N)\right) S_{i}^{z} S_{j}^{z}\bigg\} ,
\end{multline}
that corresponds to \chem{Co/Cu(100)/Co} multilayer ultrathin building with the in plane magnetization. The Hamiltonian below
\begin{multline}
H=-\sum_{\langle i,j\rangle} J_{ij}\bigg\{\left(1-\Delta_{2}(N)\right)\left(S_{i}^{x} S_{j}^{x}+S_{i}^{y} S_{j}^{y}\right) \\ + S_{i}^{z} S_{j}^{z}\bigg\},
\end{multline}
Corresponds to the \chem{Pt/Co/Cu/Co/Pt} structure with out-of-plane magnetization.

The above equations (1) and (2) are used to express the in-plane and out-of-plane magnetization configuration, but thanks to the $\Delta_{1,2}(N)$ 
parameter used in the above equations, we can specify the configuration for a specific system under study. The anisotropy parameter is presented as 
a function of the film thickness $\Delta_{1,2}(N)$ and calculated based on the approximation of experimental data presented in \cite{Huang1994, Li1992}.

$S_{i}=(S_{i}^{x} , S_{i}^{y} , S_{i}^{z} )$ is a three-dimensional unity vector at the $i$ - node of the face-centered cubic lattice.
And $J_{i,j}$ represent itself the exchange integral defining the interaction between neighboring spins $J_{1}$ takes the value $J_{1} /k_{B}T = 1$.
The exchange integral for describing the interlayer interaction is given as $J_{2}= -0.3J_{1}$. The negativity of $J_{2}$ is due to the fact that the thickness 
of nonmagnetic spacers in multilayer structures with the effect of giant magnetoresistance is chosen in such a way that the interlayer exchange interaction effectively ensures antiferromagnetism.

Below is a two-time dependence for the autocorrelation function: 
\begin{multline} 
C(t,t_w)= \biggl< \frac{1}{NL^2} \sum_{i = 1}^{NL^2}
\vec{S}_i(t) \vec{S}_i(t_w) \biggr> \\ - \vec{m}(t)\vec{m}(t_w),
\end{multline}

Angle brackets in equation (3), (4) indicate statistical averaging. In the presented study, calculations were carried out for cobalt-based films, taking 
into account the different thicknesses of the thin films under study.  Important to note that for large waiting times, the decay of the two-time autocorrelation
function calculated depending on the observation time occurs more slowly; $m$ - magnetization of the ferromagnetic film.

In Fig.~\ref{fig:image_1}.(a,b) we see the time dependence of the autocorrelation function $C(t,t_{w})$ on the observation time for different times $t_{w} = 30$ and $t_{w} = 100$ ml/s. 
Two thicknesses of ferromagnetic films were chosen for the study (the value is measured in monolayers): from $N = 3$ to $N = 9$. The study temperature was chosen to be half the 
critical temperature (where the critical temperature is measured in exchangeable integer units) to examine the behavior of the system in the low temperature region considered to be 
closer to practical applications. In these graphs, we can see the effects of aging emerging. The evolution of the system was considered for a high-temperature initial state. 
The graphs clearly demonstrate the manifestation of aging effects in these structures, namely the slowdown of correlation processes with increasing $t_w$.

The graph Fig.~\ref{fig:image_1}(a) shows the structure of \chem{Co/Cu/Co}, it can be noted that with increasing thickness of ferromagnetic films in the system under consideration, 
an increase in the aging effects is observed, i.e. e. an even greater slowdown of correlation processes. Arguments in favor of this behavior are associated with the XY-type anisotropy 
realized in \chem{Co/Cu/Co} structures, as well as with the very slow dynamics in the two-dimensional XY model, which is described not only by aging near the phase transition temperature, 
but also throughout the entire region of existence of the low-temperature phase. In the case of ultrathin films of the $XY$ type, these features of nonequilibrium behavior are preserved.

Graph Fig.~\ref{fig:image_1}(b)., \chem{Pt/Co/Cu(100)/Co/Pt} shows the Ising type structure \chem{Pt/Co/Cu(100)/Co/Pt} it can be noted that with increasing thickness of 
ferromagnetic films in the system under consideration, a weakening of the aging effects is observed, i.e. weakening of correlation processes with increasing thickness.

I would like to note a remarkable feature of $C(t,t_{w})$ Ising-type systems, that is, we observe the opposite dependence of the number of layers of the XY-system, shown in Fig.~\ref{fig:image_1}. (a,b). 
The presented results allow us to conclude that not only the initial states of the system influence, but also the choice of structural materials. The demonstrated behavior of the \chem{Co/Cu/Co} structure 
allows us to conclude that aging effects manifest themselves in a wide low-temperature region. The results obtained must be taken into account when designing spintronic devices, since the influence of such behavior makes a significant contribution to these devices.

The magnetoresistance in a multilayer film can be taken into account using the equation presented below.
\begin{equation}
\delta = \frac{(R_{\uparrow}-R_{\downarrow})^{2}}{4R_{\uparrow}R_{\downarrow}} =  \frac{(J_{\uparrow}-J_{\downarrow})^{2}}{4J_{\uparrow}J_{\downarrow}},
\end{equation}

$R_{\uparrow}$ denotes the resistance of the ferromagnetic film groups of spin-up electrons. $R_{\downarrow}$ denotes the resistance of the ferromagnetic film groups of spin-down electrons. 
We have already provided more details about the components of the formula in the article \cite{MAIN}.  Where $J_{\uparrow, \downarrow} = en_{\uparrow, \downarrow} \langle V_{\uparrow, \downarrow} \rangle $ is the current density.
$\langle V_{\uparrow, \downarrow} \rangle $ velocities of electrons with corresponding spin projections. The average electron velocity through the electron mobility and the external electric field intensity, and after that through the probability 
of electron displacement in unit time (corresponding to one Monte Carlo step per spin) from unit cell $i$ to a neighbouring unit cell in the direction of the electric field with averaging over all film unit cells.

When studying this structure, the initial conditions of the system at times $t - t_w$, $t_w \ll t_{\rm rel}$ have a great influence on the results obtained. Aging phenomena showed themselves in the dependence of the autocorrelation function $C(t,t_{w})$ on the 
characteristic time variables, such as the waiting time $t_{w}$ and the observation time $t-t_{w}$ for $t > t_{w}$ not only through $t-t_{w}$ and appear at $t \ll t_{rel}$, that $t_{rel}$ is the relaxation period. 
The time interval from sample preparation to the start of measuring its characteristics characterizes the waiting time $t_{w}$.

In this work, we first presented the results of calculations of the nonequilibrium behavior of the autocorrelation function $C(t, t_w)$, for the system under study with linear size $L = 64$, 
results are averaged over a sufficient number of runs. And we can observe the results obtained allow us to determine the nonequilibrium characteristics of the system already at times of 5000 Monte-Carlo steps per spin (MCS/s). 

It is known that a system with a smaller linear size was previously considered\cite{RefA} in the case of studying nonequilibrium phenomena in magnetic multilayer nanostructures and aging in magnetoresistance. 
For our study, an increased linear system size was chosen, however, based on the analysis of the presented results, it can be concluded that the increase in system size have a negligible impact.

It should be noted that the graphs of magnetoresistance dependence presented in the article obtained for structures in Fig.~\ref{fig:image_5} and Fig.~\ref{fig:image_7}. Have a similar behavior, but differ with respect to the value when the system reaches that the magnetoresitance reaches a plateau.

On Fig.~\ref{fig:image_5}. shows that the magnetoresistance in the \chem{Pt/Co/Cu} structure reaches a plateau over times of the order of $4000-8000$ MCS/s. The simulation results show that the magnetoresistance $\delta(t,t_w)$ depends on the 
waiting time $t_w$ as a common aging criterion and that $\delta(t,t_w)$ plateaus with asymptotic values $\delta^{\infty }(T,N)$, what depend on the type of the initial state, the thickness of the \chem{Co} films, and the temperature. It is found that the 
values of $\delta^{\infty}(T,N)$ obtained for the case of system evolution from a low-temperature initial state are in good agreement with the equilibrium values of the magnetoresistance $\delta^{(eq) }(N,T)$ . The meaning of $\delta^{\infty}(T,N)$ 
obtained for the happening of evolution from a high-temperature initial state differentiate from the equilibrium meaning of the magnetoresistance and are below these meaning.

A significant manifestation of slow dynamics is the violation of the fluctuation-dissipative theorem. When the connection between the system response function to an external disturbance $R(t,t_{w})$ and the correlation function $C(t,t_{w})$ is carried 
out through the introduction of an additional quantity $X(t,t_{w})$, called the fluctuation-dissipation ratio (FDR). In the long-term mode $(t-t_{w})\rightarrow \infty$ the dynamic susceptibility can be written as:

\begin{multline}
T_{s}\chi(t,t_{w}) = \int_{0}^{t_{w}}X(t,t')\frac{\partial C(t,t')}{\partial t'}dt' \\ 
= \int_{0}^{C(t,t_{w})}X(C)dC
\end{multline}

In equation (5), the dynamic susceptibility $\chi(t,t_{w})$ is the thermoremanent (TR) susceptibility,  which is given by the integration of the response function from $0$ to $t_{w}$ and measured after the magnetic field applied at $0$ is turned off at $t_{w}$.

Then we obtain the fluctuation-dissipation ratio as a function of the waiting time:

\begin{equation}
X(t_{w}) = \lim\limits_{C(t,t_{w})\rightarrow \infty} T_{s} \frac{\partial\chi(t,t_{w})}{\partial C(t,t_{w})} 
\end{equation}

For times $t > t_{w} >> t_{rel}$ - the relaxation time of the system, FDR establishes that $X(t,t_{w})$. The asymptotic value of the fluctuation-dissipation ratio is an important universal characteristic of nonequilibrium processes in various systems, 
which is a measure of violation of the fluctuation-dissipation theorem and is presented in the form:

\begin{equation}
X^{\infty} = \lim\limits_{t_{w} \rightarrow \infty}X(t_{w})
\end{equation}

Knowing $X(t_{w})$ for various waiting times and extrapolating $X(t_{w}\rightarrow\infty)$, we can determine the limiting FDR $X^{\infty}$. Which, according to the fluctuation-dissipation theorem, should be equal
unit $X^{\infty}=1$, which is not always true.


Violation of the fluctuation-dissipation theorem is a significant manifestation of the slow one. An important universal measure of FDT violation in various systems is the fluctuation-dissipation relationship, which can be specified in the following form:

\begin{multline} 
X^{\infty} = \lim_{t_{w}\rightarrow \infty} \lim_{t \rightarrow \infty} X(t,t_{w}) = \\
 \lim_{t_{w}\rightarrow \infty} \lim_{C \rightarrow \infty} T \frac{\partial X(t,t_{w})}{\partial C(t,t_{w})}
\end{multline}

\begin{figure}[!th]
\centering
{\includegraphics[width=0.4\textwidth]{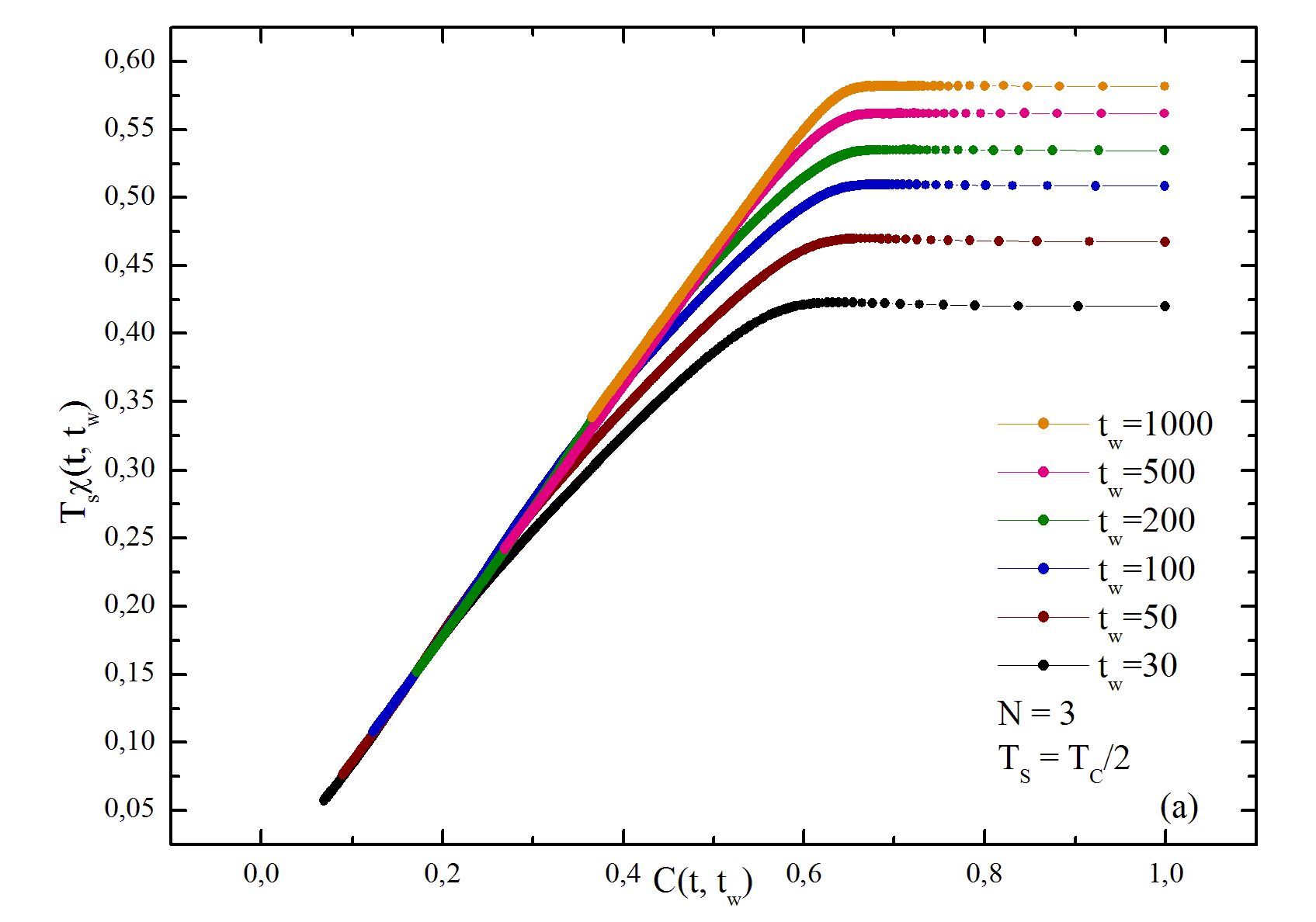}%
} 
{\includegraphics[width=0.4\textwidth]{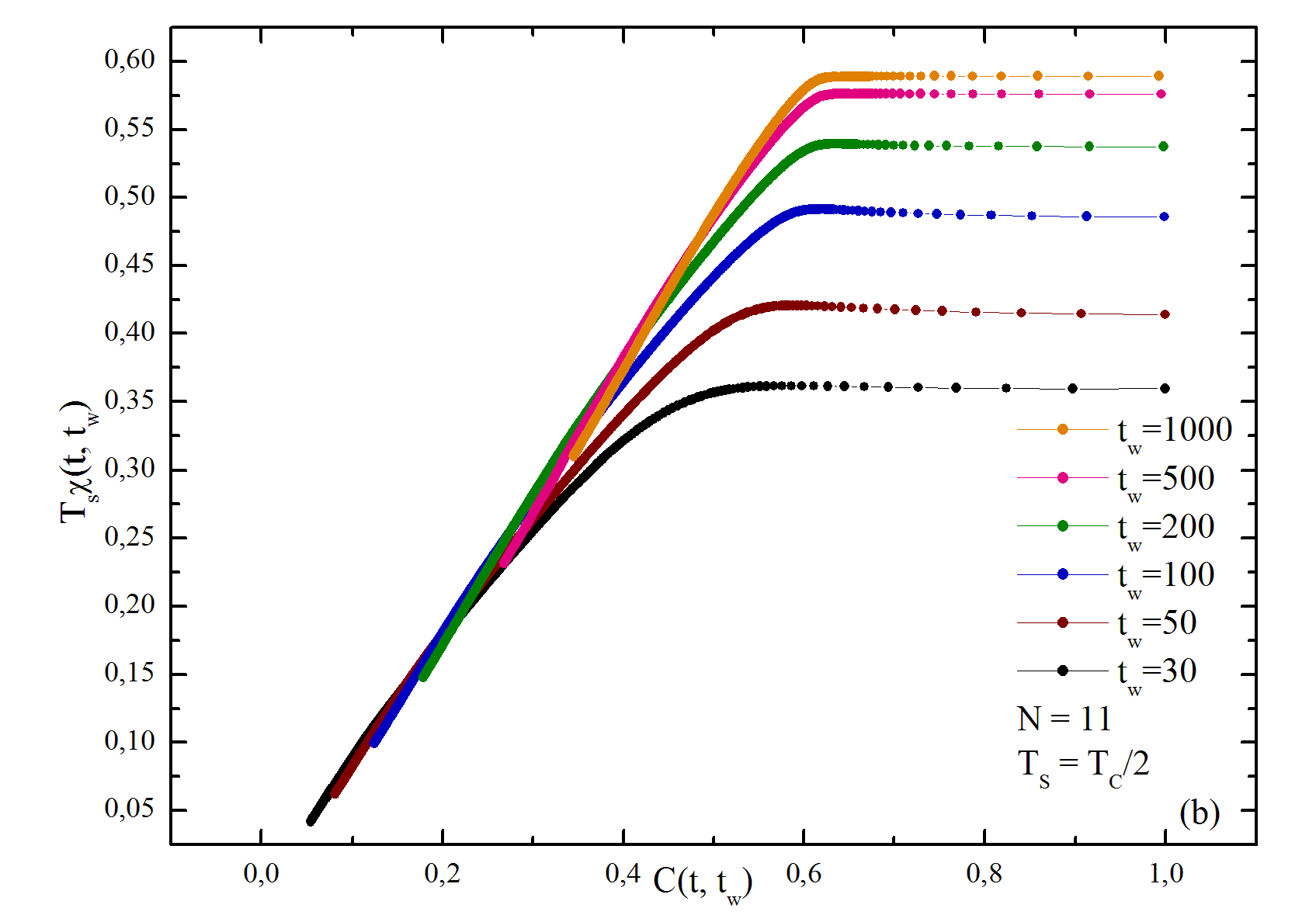}%
} 
\caption{\label{fig:image_6} Parametric dependence of the dynamic susceptibility $T_{s} \chi(t,t_{w})$ on the autocorrelation function $C(t,t_{w})$ for different waiting times $t_{w}$, at freezing temperature $T_ {s} = T_{c}/2$ during evolution from a 
high-temperature initial state for films with a thickness of $N = 3$ and $N=11$ for \chem{Co/Cu/Co}.}
\end{figure}

\begin{figure}[!th]
\centering
{\includegraphics[width=0.4\textwidth]{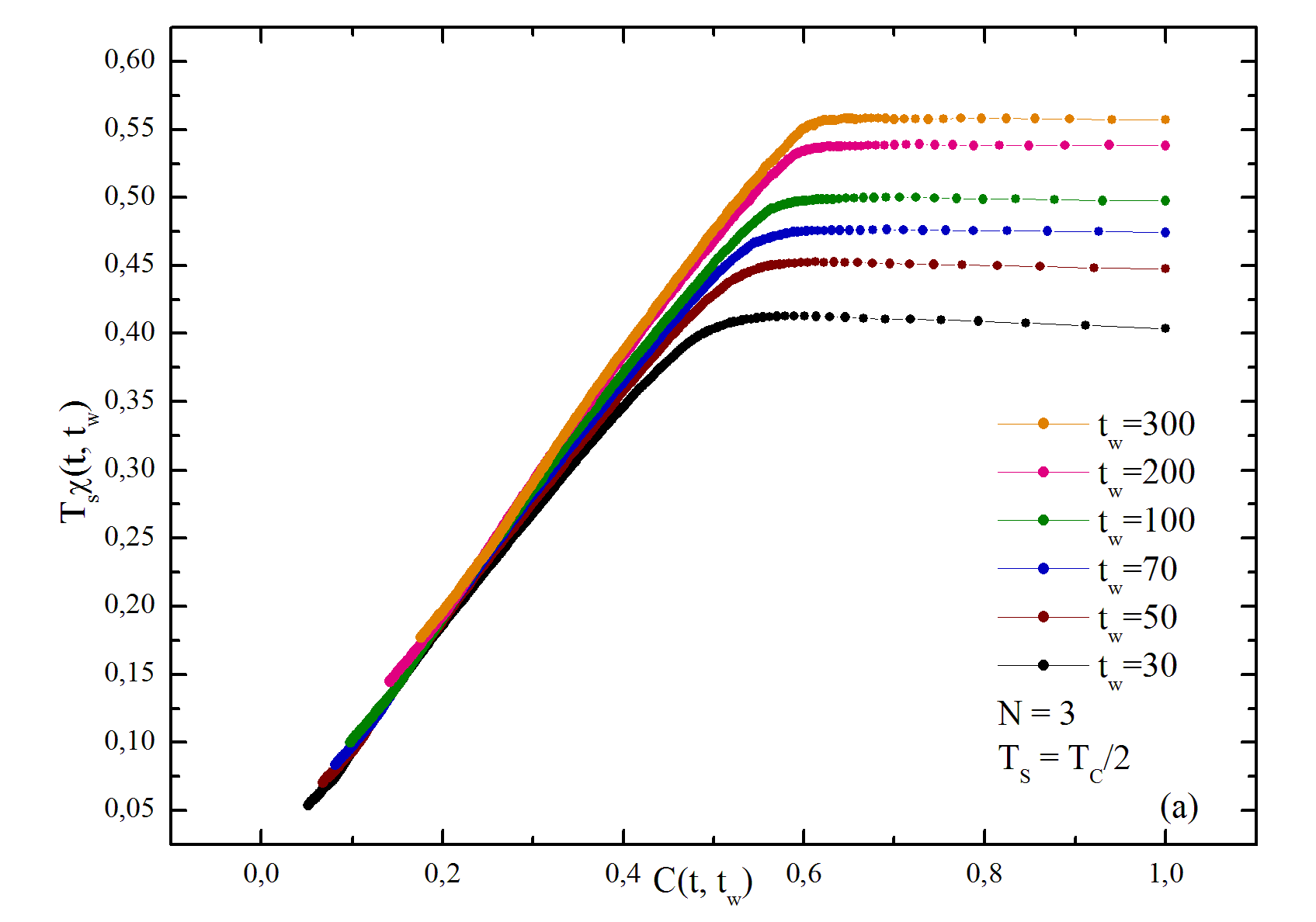}%
} 
{\includegraphics[width=0.4\textwidth]{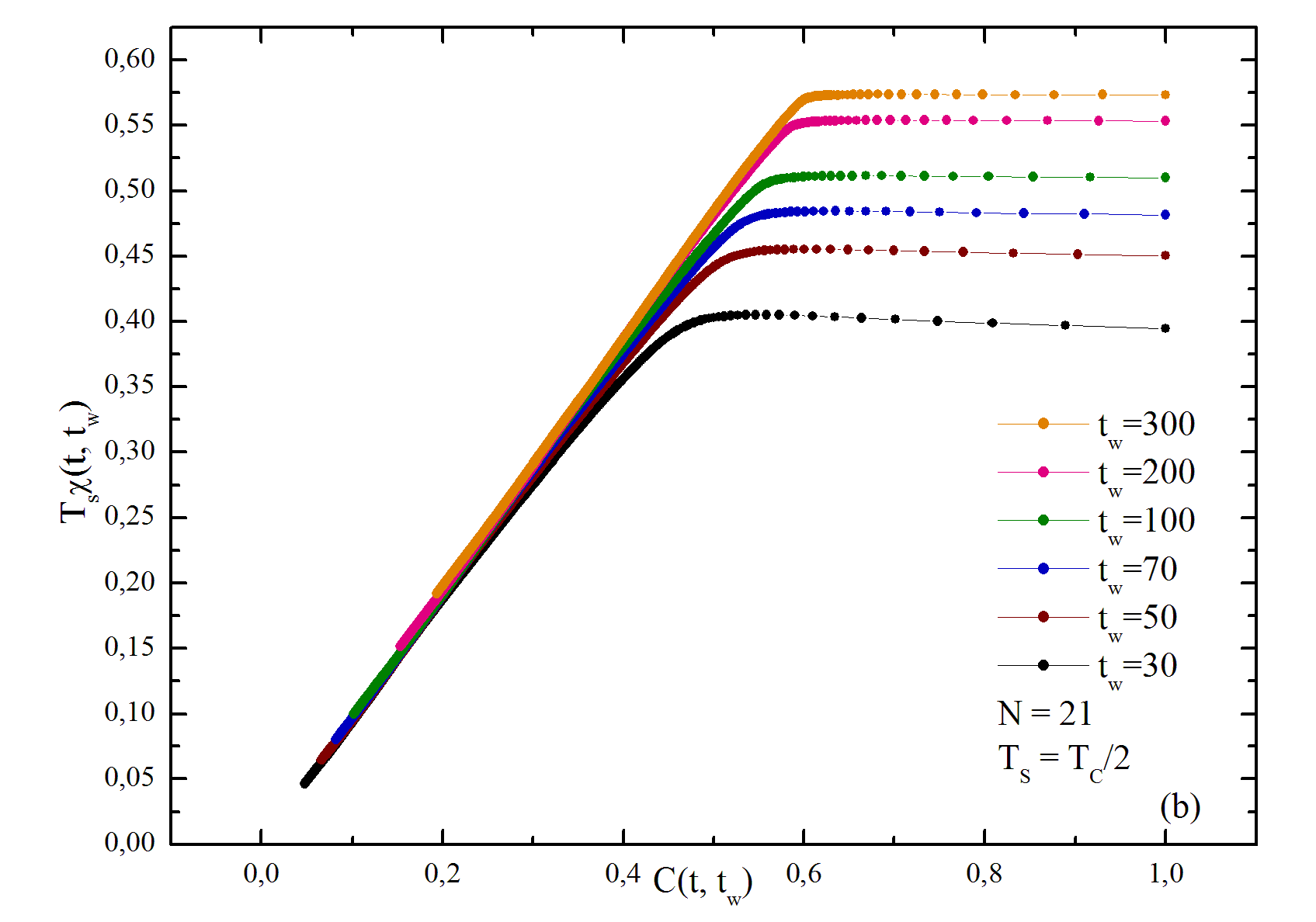}%
} 
\caption{\label{fig:graph_5} Parametric dependence of the dynamic susceptibility $T_{s} \chi(t,t_{w})$ on the autocorrelation function $C(t,t_{w})$ for different waiting times $t_{w}$, at freezing temperature $T_ {s} = T_{c}/2$ during evolution from a 
high-temperature initial state for films with of $N = 3$ and $N=21$ for \chem{Pt/Co/Cu/Co/Pt}.}
\end{figure}

Next, we clarify the general properties of the quantity $X^{\infty}$ and its dependence on the freezing temperature $T_{s}$ of the system. Thus, for the state of the system with the temperature $T_{s} > T_{c}$, it is established as a consequence of FDT that $X^{\infty} (T_{s} > T_{c}) = 1$. On the other hand, based on the general scaling arguments in the work \cite{Ref2017Prud}, it is shown that for the low-temperature ordered phase with $T_{s} < T_{c}$, the quantity $X^{\infty} (T_{s} < T_{c}) = 0$. It is expected that these results do not depend on the specific properties of individual systems. However, in the case of $T_{s} = T_{c}$, there are no general arguments establishing the value of $X^{\infty}(T_{c})$, so it must be determined for each individual statistical model. The values $X^\infty (T_s = T_c )$ depends on the specific properties of the model and its spatial dimension $d$ (Table 1 from  \cite{Ref2017Prud}). However, in the papers \cite{Ref25, Ref27} it is claimed on the basis of scaling arguments that at the critical temperature the limiting FDT $X^{\infty} (T_{s} = T_{c})$ should be a universal quantity related to the universality class of the critical dynamics of the model.

To determine the limiting values of the FDR for systems with different thicknesses of ferromagnetic films $N = 3, N = 11$ and
at different temperatures, a parametric dependence of the dynamic susceptibility on the autocorrelation function was
constructed. The graphs of this parametric dependence allow us to determine the values of the FDR $X(t_{w})$ for each
waiting time $t_{w}$ using the asymptotic curvature. By applying linear approximation and extrapolation procedures 
$X^{\infty}(t_{w} \rightarrow \infty )$ to the obtained values, it is possible to determine the limiting values of the FDR.

Below is an analysis of the behavior of the dependence of the dynamic susceptibility $T_{s} \chi(t,t_{w})$ on the autocorrelation function $C(t,t_{w})$ for different waiting times $t_{w}$ at freezing temperature $T_{s} = T_{c}/2$.

On Fig.~\ref{fig:image_6} shows the parametric dependence of the dynamic susceptibility; calculations of the system were carried out during the evolution from a high-temperature initial state; thicknesses of $N = 3$ and $N = 11$ 
were chosen for the calculations. To calculate the dynamic susceptibility, we used the anisotropic Heisenberg model taking into account the influence of anisotropy; for films with a thickness of N = 3, we used the value $\Delta (N = 3) = 0.665$ and $
\Delta (N = 11) = 0.22$, which indicates the behavior of the system is close to the XY model. The results obtained make it possible to determine the values of $X (t_ {w})$ from the asymptotic curvature for each waiting time $t_ {w}$.

Below is an analysis of the behavior of the dependence of the dynamic susceptibility $T_{s} \chi(t,t_{w})$ on the autocorrelation function $C(t,t_{w})$ for different waiting times $ t_{w}$ at freezing temperature $T_{s} = T_{c}/2$.

\begin{figure}[!th]
\centering
{\includegraphics[width=0.44\textwidth]{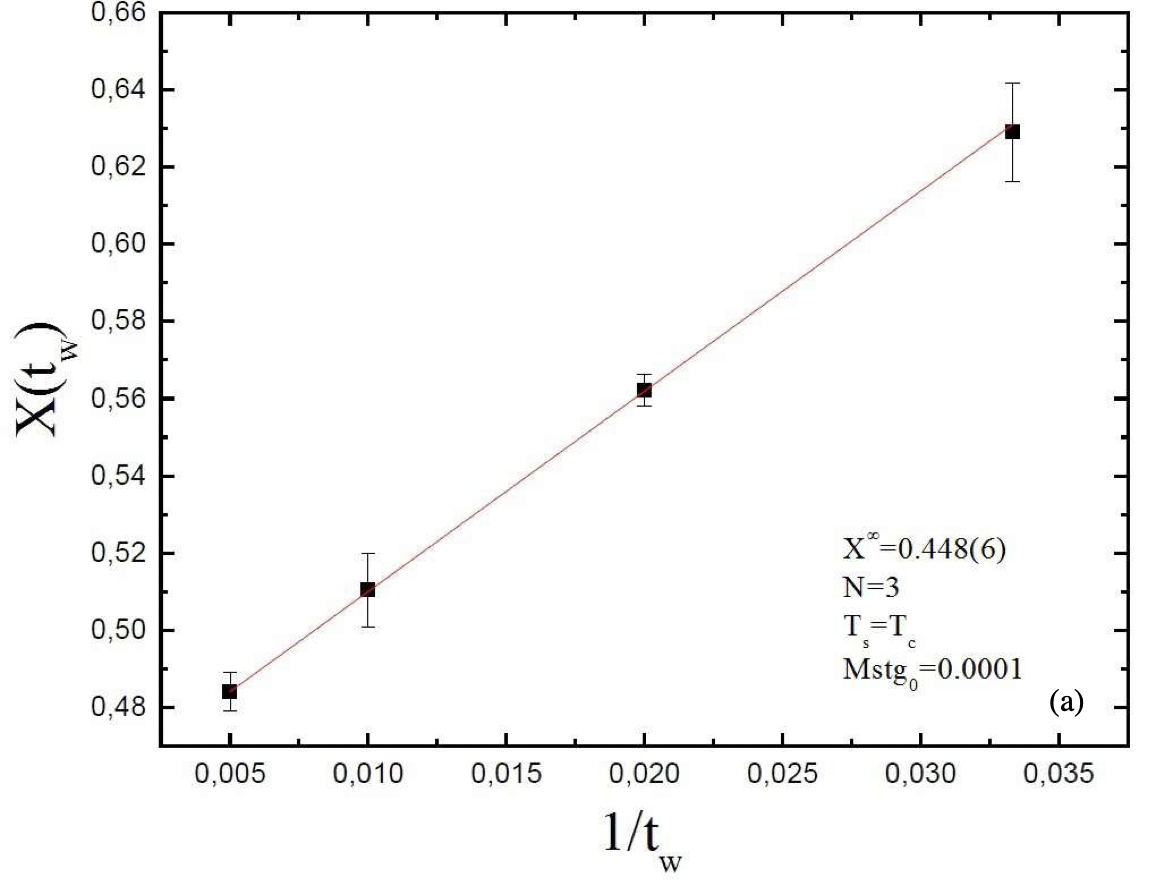}%
} 
{\includegraphics[width=0.42\textwidth]{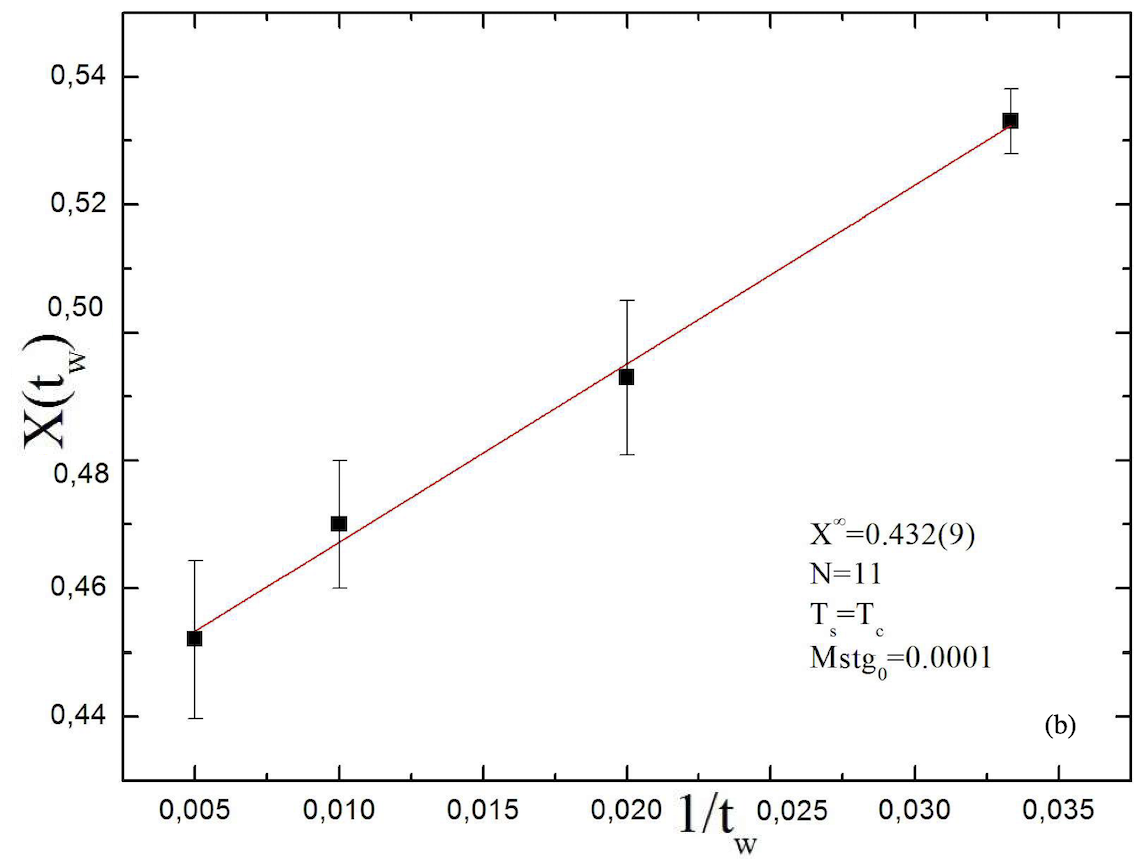}%
} 
\caption{\label{fig:graph_6} The linear dependence $X(t_{w})$ at $1/t_{w}$ for at freezing temperature $T_ {s} = T_{c}$ for films with a thickness of $N = 3$ and $N = 11$ for \chem{Co/Cu/Co} system.}
\end{figure}

Fig.~\ref{fig:graph_6} shows the results of the procedure for determining the FDT. Regarding the question of determining FDT. Linear approximation of the dependence $X(t,t_{w})$ at $t_{w} / (t - t_{w}) \rightarrow 0$ allowed us to determine the values $X(t_{w})$ for each $t_{w}$. The extrapolation $X(t_{w} \rightarrow \infty)$ was applied to the obtained values $X(t_{w})$, which allowed us to determine the desired limiting FDT. By applying the above procedure we determined $X^{\infty} = 0.448(6)$ for the magnetic structure with $N=3$. 
This value was obtained in good agreement with the values obtained from the two-dimensional XY model ($X^{\infty} = 0.444(26)$) at temperature $T_{BKT}$, thus it is shown that this is ensured influence of magnetic anisotropy 
of orienting magnetization in the plane. This indicates two-dimensional effects in the behavior of structures with ultrathin films.

For a structure with a thickness $N=11$, the limiting FDR is calculated during the evolution from a high-temperature initial state. The obtained value $X^{\infty} = 0.432(9)$ within the error agrees well with the value of the limiting FDR for the three-dimensional XY model.

\begin{figure}[!th]
\centering
{\includegraphics[width=0.44\textwidth]{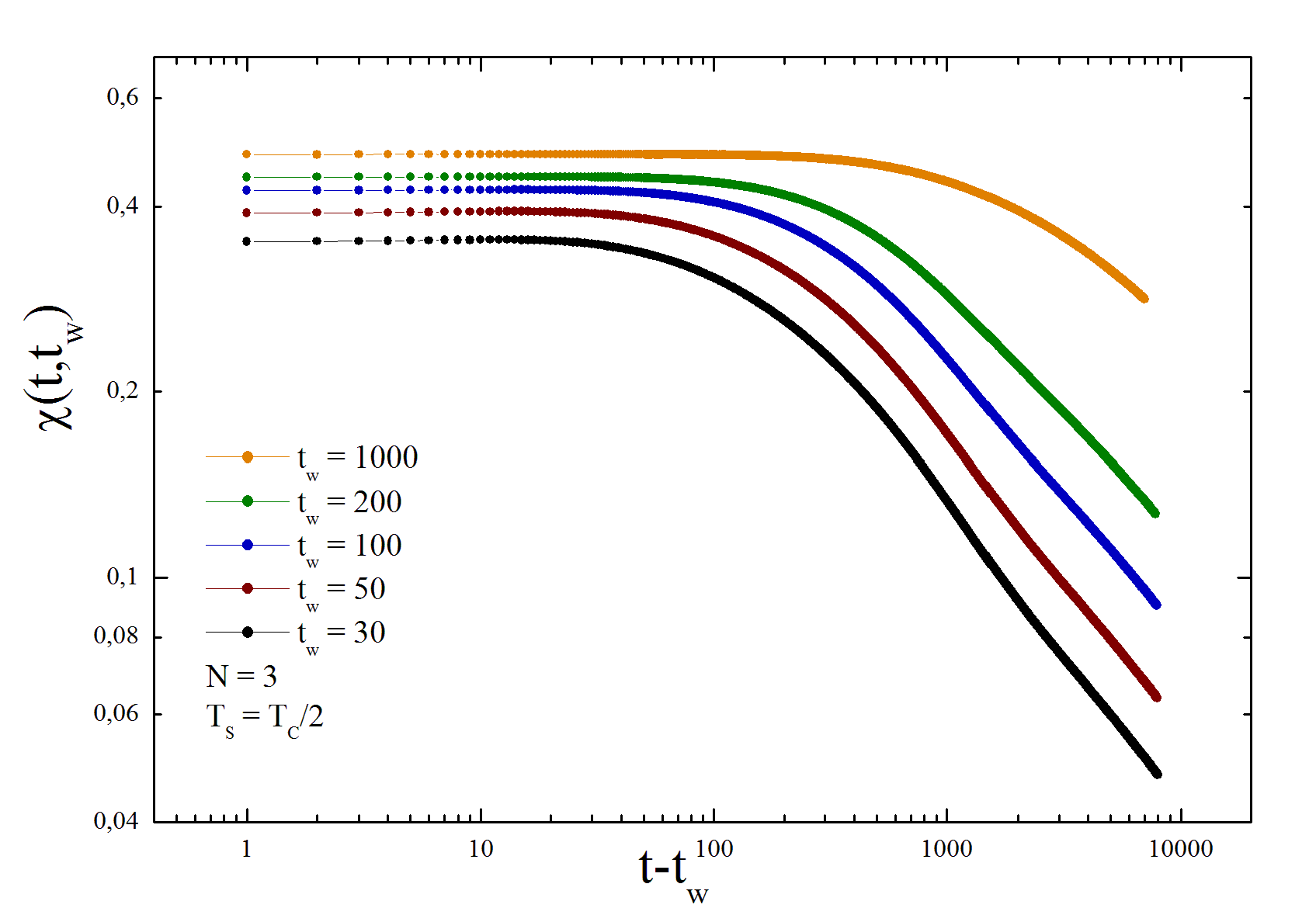}%
} 
\caption{\label{fig:N3_CoCu_Tc2_Chitw} Dynamic susceptibility $\chi(t,t_{w})$ for the system \chem{Co/Cu/Co} at $N=3$, for temperature $T_{s} = T_{c}/2$ at different values $t_{w}$  during evolution from a high-temperature initial state.}
\end{figure}

Multilayer magnetic nanostructures as systems with slow dynamics are characterized by the manifestation of aging effects for dynamic susceptibility from the waiting time and observation time, which are characterized by a slowdown in the correlation and relaxation processes. Therefore, it can be argued that these functions are important characteristics in the study of nonequilibrium critical dynamics of the system. As shown in Fig.~\ref{fig:N3_CoCu_Tc2_Chitw}, we observe the manifestation of size effects for dynamic susceptibility, which are characterized by a slowdown in time decay with increasing film thickness in the low-temperature region. 

\section{CONCLUSION}
We used in Monte-Carlo simulations to compute of the nonequilibrium behavior of \chem{Pt/Co/Cu(100)/Co/Pt} and \chem{Co/Cu(100)/Co} multilayer nanostructures by using anisotropic Heisenberg model to describe the magnetic properties of ultrathin films based of
\chem{Co}. This arises due to an increase in the characteristic spin-spin correlation length in magnetic ultrathin structures. The manifestation of aging effects is expressed in a slowdown in system correlation time with an increase in waiting time.

As for the results of the dependence of the behavior of the system on the thickness of the films in the case of the \chem{Pt/Co/Cu}  system at temperatures $T_c(N)/2$, we can observe a decrease in slowing down  $C(t,t_w)$ with increasing thickness $N$.

It is important to highlight that aging effects manifest themselves in the low-temperature region ($T_ {s} = T_{c}/2$) in the case of not only the dependence of the autocorrelation function, but also in the case of dynamic susceptibility. This behavior is also confirmed by the
obtained values of the $\delta$ parameter.

It is found that the magnetoresistance achieves a tableland in the asymptotic long-term mode with $\delta^\infty(N,T)$ values, which rely on the kind of initial state, cobalt film thickness, temperature, and type of magnetic anisotropy in such structures.

An important addition to our results are calculations of the parametric dependence of the dynamic susceptibility $T_{s}$ by analyzing which we can also observe the influence of aging effects. 

The manifestation of aging effects in a wide low-temperature region, in addition to the region near the critical temperature, affects devices that belong to spintronic devices. Therefore, the obtained results presented in this study should be taken into account when designing spintronic devices, since the influence of such behavior makes a significant contribution to these devices.

\acknowledgements

This research was supported by RFBR projects for young PhD students 20-32-90226 (M.M.B.), 19-32-90261 (V.O.B.), by Russian Science Foundation, project 23-22-00093 (M.V.M, V.V.P.), by the Ministry of Science and Higher Education of the Russian Federation within the government assignment for Boreskov Institute of Catalysis, project FWUR-2024-0039 (P.V.P.). Marina M. Boldyreva express gratitude for the to support the research of Prof. Tatsuki Oda from Kanazawa University (Japan) and the Council for Grants of the President of the Russian Federation for the to support the research. 

The studies were carried out using the resources of the Center for Shared Use of Scientific Equipment “Center for Processing and Storage of Scientific Data of the Far Eastern Branch of the Russian Academy of Sciences”, which is supported by the Ministry of Science and Higher Education of the Russian Federation (project No. 075-15-2021-663) \cite{bib:VC_DVO_RAN}.

\end{document}